\documentclass[11pt,a4paper,twoside]{article}
\usepackage{amsmath}
\usepackage{amsfonts}
\usepackage{rotating}
\usepackage{graphicx}
\usepackage[T1]{fontenc} 

\usepackage[english]{babel}
\usepackage{pdflscape}
\usepackage{float}

\numberwithin{equation}{section}
\begin{document}
\vspace{1cm}

\noindent
{\bf
{\large Duality and Dimensional Reduction of 5D BF Theory
}}

\vspace{.5cm}
\hrule

\vspace{1cm}

\noindent

\begin{center}{\bf  A. Amoretti, A. Blasi, G. Caruso, N. Maggiore and N. Magnoli}
\footnote{$name$@ge.infn.it}\end{center}
\begin{center}{\footnotesize {\it
 Dipartimento di Fisica, Universit\`a di Genova \\
via Dodecaneso 33, I-16146 Genova -- Italy \\and \\INFN, Sezione di
Genova 
} }
\end{center}

\vspace{1cm}
\noindent

{\tt Abstract~:}
A planar boundary introduced {\it \`a la} Symanzik in the 5D topological BF theory, with the only requirement of locality and power counting, allows to uniquely determine a gauge invariant, non topological 4D Lagrangian. The boundary condition on the bulk fields is interpreted as a duality relation for the boundary fields, in analogy with the fermionization duality which holds in the 3D case. This suggests that the 4D degrees of freedom might be fermionic, although starting from a bosonic bulk theory. The method we propose to dimensionally reduce a Quantum Field Theory and to identify the resulting degrees of freedom can be applied to a generic spacetime dimension.

\vfill\noindent
{\footnotesize {\tt Keywords:}
Quantum Field Theories with Boundary,
Topological Quantum Field Theories, 
Canonical Quantization,
Gauge Symmetry.
\\
{\tt PACS Nos:} 
03.70.+k	Theory of quantized fields
11.10.-z	Field theory
11.15.-q	Gauge field theories
11.10.Kk	Field theories in dimensions other than four
11.15.Yc	Chern-Simons gauge theory
04.60.Ds	Canonical quantization
11.25.Mj	Compactification and four-dimensional models.
}
\newpage
%*********************************************************************************************
%%%%%%%%%% end of title page %%%%%%%%%%%%%%%%%%%%%%%%%%%%%%
\begin{small}
%\tableofcontents
\end{small}

\setcounter{footnote}{0}

%*********************************************************************************************
\section{Introduction}
%*********************************************************************************************

This paper has a twofold purpose: the first one is to describe how the 5D abelian BF model defined on a manifold with a planar boundary uniquely determines a 4D (non-topological) theory. The second is to propose the BF model as the generator of a duality relation which leads to the ``fermionization'' of the boundary degrees of freedom, as discussed in \cite{Aratyn:1984jz}.

This kind of duality is extremely useful in condensed matter physics since it provides a tool of systematically deriving effective hydrodynamical field theories for generic topological phases \cite{Chan:2012nb}.

In recent years the study of  TQFT on a manifold with a boundary acquired a remarkable physical relevance, since these models constitute effective field theories which describe  some aspects of  new materials, the so called Topological Insulators \cite{Hasan:2010xy}. In fact, in \cite{Cho:2012tx,Santos:2011bf} it was shown that the 3D BF model represents an effective field theory for Quantum Spin Hall Systems, while, in \cite{Cho:2012tx} the 4D BF model with boundary was suggested as an effective field theory for 3D Topological Insulators.

From the theoretical point of view the study of TQFT with a boundary is interesting for on the edge they acquire local observables. Among the TQFT, the BF models have the peculiarity of being defined for any spacetime dimension D, and hence they allow to investigate what happens on a boundary of any (D-1) dimension.

Now two questions naturally arise: does the BF model uniquely determine the dynamics on its boundary and do we always find a duality relation which yields the ``fermionization'' of the residual bosonic degrees of freedom ? 

The aim of this paper is to contribute to answer these two questions by analyzing the abelian 5D BF model with a planar boundary.

Our approach to the introduction of a boundary term in a QFT follows the lines proposed by Symanzik in \cite{Symanzik:1981wd}, and in particular the idea of ``separability'' $i.e.$ the request that there is no propagation across the boundary plane.

As a result we find that the Symanzik method applied to 5D BF theory leads to a 4D model whose Lagrangian is uniquely identified. One of the key ingredients needed for this identification is the 4D duality relation which is also responsible of the ``fermionization'' procedure.

This result, and the analogous ones already obtained for the 3D and 4D BF models \cite{Blasi:2011pf,Amoretti:2012kb}, strongly suggests that the fermionization on the boundary of the purely bosonic degrees of freedom, is a common feature of all BF models. \\

The paper is organized as follows. In section 2 the classical 5D BF model is described, while the boundary is introduced in section 3 and the boundary conditions for the fields of the bulk theory are discussed in section 4. Section 5 and section 6 are the heart of the paper. Here we derive the scalar-vector duality previously discussed and we deduce the unique 4D theory living on the boundary of the 5D model.
%*********************************************************************************************
\newpage\section{The classical theory}
%*********************************************************************************************

We consider the 5D abelian BF model, which describes the interaction between the antisymmetric rank 3 tensor $B_{\mu \nu \rho}(x)$ and the gauge field $A_{\mu}(x)$, defined on the flat 5D Minkowski spacetime with metric $g_{\mu\nu}\equiv diag(-1,1,1,1,1)$. The action of the model is:
\begin{equation}
\mathcal S_{BF}=\frac{1}{2}\int_M{d^5x\,\epsilon^{\mu\nu\rho\sigma\tau}F_{\mu\nu}B_{\rho\sigma\tau}},
\label{bf}
\end{equation}
where $F_{\mu\nu} (x)\equiv \partial_\mu A_\nu(x)-\partial_\nu A_\mu(x)$ is the field strength  for the gauge field $A_\mu(x)$.
The action \eqref{bf} is invariant under the symmetry
\begin{eqnarray}
\delta^{(1)} A_\mu &=& \partial_\mu \theta
\\
\delta^{(1)} B_{\mu\nu\rho}&=&0,
\end{eqnarray}
which is the usual gauge transformation, and under
\begin{eqnarray}
\delta^{(2)} A_\mu &=&0
\\
\delta^{(2)} B_{\mu\nu\rho}&=&\partial_\mu \phi_{\nu\rho} + \mbox{ cyclic permutations},
\end{eqnarray}
where $\theta(x)$ and $\phi_{\mu \nu}(x)$ are local gauge parameters.

It is convenient to adopt axial gauge choices for the fields $A_\mu(x)$ and $B_{\mu\nu\rho}(x)$:
\begin{eqnarray}
A_4&=&0\label{gaugecond1}
\\
B_{ij4}&=&0\label{gaugecond2},
\end{eqnarray}
where $i=\{0,1,2,3\}$ from now on. In order to implement this choice, we add to the action \eqref{bf} the gauge-fixing term:
\begin{equation}
\label{gf}
\mathcal S_{gf}=\int_M{d^5x\,(b A_4  + d^{ij}B_{ij4}}),
\end{equation}
where $b(x)$ and $d^{ij}(x)$ are Lagrange multipliers.

We did not introduce the Faddeev-Popov ghost fields since, as usual, in the abelian case they are decoupled. 

The axial gauge is particularly convenient to study Quantum Field Theories (QFTs) with boundary, and the fact that the gauge conditions \eqref{gaugecond1} and \eqref{gaugecond2} are not covariant is not a problem, since we are about to introduce the planar boundary $x_4=0$, which $per\ se$ breaks the covariance of the theory.

Summarizing, the classical action is given by:
\begin{multline}
\Gamma_c[J_{\Phi}]=\int_M{d^5x\,\{\epsilon^{ijkl}[3\partial_i A_j B_{kl4}}+(\partial_4 A_i - \partial_i A_4)B_{jkl}]+ A_4 b + d^{ij}B_{ij4} +
\\
+ J_i A^i + J_4 A^4 + J_{ijk}B^{ijk} +  J_{ij4}B^{ij4}+J_bb + J_{d ij}d^{ij}\},
\label{stot}
\end{multline}
where $J_{\Phi}(x)$ are the external sources coupled to the fields $\Phi(x)$.\\
Consequently, the bulk EOM derived from the classical action \eqref{stot} are:
\begin{eqnarray}
\epsilon^{ijkl}\left(3\partial_j B_{kl4}- \partial_4 B_{jkl}\right)+J^i &=&0 \label{eqm1}
\\
\epsilon^{ijkl}\left(\partial_4A_l-\partial_lA_4\right)+J^{ijk} &=&0
\label{eqm2}\\
 \epsilon^{ijkl}\partial_i B_{jkl}+J^4+b &=&0
\label{eqm3}\\
 3\epsilon^{ijkl}A_l+J^{ij4}+d^{ij} &=&0
\label{eqm4}\\
 A_4+J_b &=&0
\label{eqm5}\\
 B^{ij4}+J^{ij}_d &=&0.
\label{eqm6}\end{eqnarray}
It is well known that the gauge-fixing term \eqref{gf} does not completely fix the gauge \cite{Bassetto:1991ue}. The gauge-fixed action \eqref{stot} is still invariant under gauge transformations in the directions orthogonal to $x_4$, and the residual gauge invariance is functionally expressed by two local Ward Identities (WI), (one for each symmetry $\delta^{(1)}$ and $\delta^{(2)}$):
\begin{eqnarray}
W(x) \Gamma_c[J_{\Phi}]&=&\partial_i J^i +\partial_4 J^4+\partial_4 b= 0\label{wi1}
\\
W^{ij}(x) \Gamma_c[J_{\Phi}]&=&-3\partial_k J^{ijk}+\partial_4J^{ij4}+\partial_4d^{ij}=0,\label{wi2}
\end{eqnarray}
as it can be checked directly from the EOM \eqref{eqm1}-\eqref{eqm6}.

%*********************************************************************************************
\section{The boundary}
%*********************************************************************************************

To introduce a boundary in the theory, we adopt the method described by Symanzik in \cite{Symanzik:1981wd}, which allows to describe a boundary in a QFT in a very general and smooth way. It basically consists in adding to the bulk action \eqref{stot} the most general boundary term compatible with the fundamental QFT principles of locality and power-counting, and then computing, for the modified theory, the propagators, with the constraint that the propagators between points lying on opposite side of the boundary $x_4=0$, vanish. This condition is known as ``separability'', and, although its formulation is quite simple, it might render the explicit calculation of the propagators very difficult. However, this task can be avoided if, as in our case, one is interested only in the physics on the boundary. In fact, an improved version of the Symanzik's method has been introduced in \cite{Amoretti:2012kb,Blasi:2010gw}, which leads directly to a ``separated'' theory with boundary, without the need of calculating explicitly the propagators. It will turn out that the boundary dynamics is completely determined by the WI \eqref{wi1} and \eqref{wi2}, broken by the most general (linear) separating boundary term.

Following the steps described in \cite{Amoretti:2012kb,Blasi:2010gw}, we introduce in the theory the planar boundary $x_4=0$ by adding to \eqref{stot} the most general boundary Lagrangian, compatible with locality and power counting:
\begin{multline}
\mathcal L_{BD}=\delta(x_4)\left[a_1 \epsilon^{ijkl}A_i B_{jkl}+\frac{a_2}{4}F_{ij}F^{ij}+\frac{a_3}{4}G_{ij}G^{ij}\right.+\label{lbd}
\\
\left.-\frac{a_4}{2} m^2 A_i A^i
+ a_5 b + \frac{a_6}{2} d_{ij}A^i A^j\right],
\end{multline}
where $G_{ij}(x)=\partial_i A_j(x)+\partial_j A_i(x)$ and $a_{1,...,6}$ are constant parameters which we shall determine later. The massive parameter $m$ has been introduced to render $a_4$ dimensionless.

Writing $\mathcal L_{BD}$, we limited ourselves to quadratic terms only. The reason for this strong constraint resides in a non-renormalization theorem \cite{Becchi:1974jw}, which guarantees that symmetry breaking terms which are only linear in the quantum fields, do not acquire quantum corrections, and hence are acceptable. It is clear that $\mathcal L_{BD}$ \eqref{lbd} induces harmless linear terms in the r.h.s. of the WI \eqref{wi1} and \eqref{wi2}.

Since the separability condition completely decouples the right and the left side of the boundary, we can focus our attention only to the half space $x_4\geq 0$, the other side being obtained by a parity transformation.

The boundary Lagrangian \eqref{lbd} modifies the EOM \eqref{eqm1}-\eqref{eqm6} as follows:
\begin{eqnarray}
\epsilon^{ijkl}\left[3\partial_j B_{kl4}- \partial_4 B_{jkl}\right]+J^i &=&
 -\delta(x_4)\left[a_1 \tilde B^{i+}+a_2\partial_j F^{ij+} -a_3 \partial_j G^{ij+}\right.
\nonumber\\&&
\left.-a_4m^2A^{i+} + a_6 d_{ij}^+A^{j+}\right]
\label{beom1}\\
\epsilon^{ijkl}\left[\partial_4A_l-\partial_lA_4\right]+J^{ijk}
&=&
-\delta(x_4)a_1\epsilon^{ijkl}A^+_l
\label{beom2}\\
\epsilon^{ijkl}\partial_i B_{jkl}+J^4+b&=&0
\label{beom3}\\
 3\epsilon^{ijkl}A_l+J^{ij4}+d^{ij}&=&0
\label{beom4}\\
A_4+J_b&=&-\delta(x_4)a_5
\label{beom5}\\
 B^{ij4}+J^{ij}_d&=&-\delta(x_4)\frac{a_6}{2} A^{i+ }A^{j+},
\label{beom6}\end{eqnarray}
where the superscript ``+'' indicates the field on the r.h.s. of the plane $x_4=0$.

The boundary term ${\cal L}_{BD}$ modifies the Ward identities \eqref{wi1} and \eqref{wi2} as well, by means of linear breakings:
\begin{eqnarray}
W(x) \Gamma_c[J_{\Phi}] 
&=& 
-\delta(x_4)[a_1 \partial_i\tilde B^{i+}-a_3\partial_i\partial_j G^{ij+}
\nonumber\\
&& -a_4m^2\partial_i A^{i+} + a_6\partial_i( d_{ij}^+A^{j+})]
\label{bwi1}\\
W^{ij}(x) \Gamma_c[J_{\Phi}]
&=&
\delta(x_4)3a_1\epsilon^{ijkl}\partial_k A^+_l,
\label{bwi2}
\end{eqnarray}
where we introduced the 4D vector $\tilde B^{i}(x)$, dual of $B_{ijk}(x)$:
\begin{equation}
\tilde B^{i} \equiv \epsilon^{ijkl} B_{jkl}, 
\label{bdual}\end{equation}
and, consequently, $J^{ijk}(x)\equiv\epsilon^{ijkl}\tilde J_{l}(x)$. 

Once integrated, the previous WI read:
\begin{eqnarray}
\int{dx_4\,\partial_i J^i} &=& -a_1 \partial_i\tilde B^{i+}+a_3\partial_i\partial_j G^{ij+} +a_4m^2\partial_i A^{i+}\nonumber\\
&&  + a_6\partial_i( d_{ij}^+A^{j+})\label{wii1}
\\
\int{dx_4\, \epsilon^{ijkl}\partial_k \tilde{J}_l} &=&-a_1\epsilon^{ijkl}\partial_k A^+_l.\label{wii2}
\end{eqnarray}

%*********************************************************************************************
\section{The boundary conditions}
%*********************************************************************************************

In the original formulation of the Symanzik's method \cite{Symanzik:1981wd}, the boundary conditions are obtained by computing the propagators for the theory with boundary and by imposing on them the separability condition. As anticipated, in this paper we are interested in the dynamics on the boundary of the theory and, as discussed in \cite{Amoretti:2012kb,Blasi:2010gw}, it is possible to get a ``good'', separating boundary term without computing explicitly the propagators.

The first step is to obtain the boundary conditions for the bulk fields. To get them, we integrate the broken EOM \eqref{beom1}-\eqref{beom6} with respect to $x_4$ in the infinitesimal interval $[-\varepsilon,\varepsilon]$, and we evaluate the expression obtained in the weak limit $\varepsilon \to 0$. This yields:
\begin{eqnarray}
 (a_1-1) \tilde B^{i+} &=& -a_2\partial_j F^{ij+} +a_3 \partial_j G^{ij+}+a_4m^2A^{i+}+ a_6 d_{ij}^+A^{j+}
\label{bordoo1}\\
(1+a_1)A_l^+&=&0
\label{bordoo2}\\
 a_5&=&0
\label{bordoo3}\\
 a_6 A^{i+}A^{j+}&=&0.\label{bordoo4}
\end{eqnarray}
We have to find out the set of parameters $a_i$ and the boundary conditions for the fields at the boundary $\left.A^{+}_i(x)\right|_{x_4=0^+}\equiv A^{+}_i(X)$ and 
$\left.\tilde{B}^{i+}(x)\right|_{x_4=0^+}\equiv \tilde{B}^{i+}(X)$ (where $X\equiv (x_0,x_1,x_2,x_3)$), which satisfy the equations \eqref{bordoo1}-\eqref{bordoo4}, with the constraint that the r.h.s. of the broken WI \eqref{wii1} and \eqref{wii2} are different from zero. Otherwise, an inconsistency would appear when deriving with respect to the external sources $J$. Under this respect, as already remarked in \cite{Amoretti:2012kb}, the boundary term plays the role of gauge fixing for the residual gauge invariance of the theory with boundary, since its presence is necessary to compute the propagators.

With the above prescriptions, there is only one solution of the system \eqref{bordoo1}-\eqref{bordoo4}:
\begin{eqnarray}
a_1&=&-1
\\
a_5&=&a_6=0\\
2 \tilde B^{i+}&=&a_2\partial_j F^{ij+} -a_3 \partial_j G^{ij+}-a_4m^2A^{i+}.\label{bbordo}
\end{eqnarray}
This solution corresponds to the unique couple of broken WI:
\begin{eqnarray}
\int{dx_4\,\partial_i J^i} &=& - \partial_i\tilde B^{i+}\label{w1}
\\
\int{dx_4\,\epsilon^{ijkl}\partial_k \tilde{J}_l} &=& \epsilon^{ijkl}\partial_k A^+_l.\label{w2}
\end{eqnarray}
We shall see in the next section that the broken WI \eqref{w1} and \eqref{w2} lead to a kind of electromagnetic structure on the boundary. It will turn out that the physics on the boundary is insensitive to the parameter $a_2$ (and to the relative boundary term), and a nonvanishing $a_3$ would lead to unphysical solutions on the boundary, and hence it will be necessarily put equal to zero, leaving $a_4$ as the only parameter on which the solution depends.

%*********************************************************************************************
\section{4D scalar -- vector duality}
%*********************************************************************************************

Following our modified Symanzik's method, we have been able to construct the most general separating boundary term, and we found the broken WI \eqref{w1} and \eqref{w2}, which describe the residual broken gauge invariance on the 4D boundary, whose spacetime coordinates are collectively denoted by $X=(x_0,x_\alpha)$, $\alpha=1,2,3$. Let us consider the WI \eqref{w1} and \eqref{w2} at vanishing external sources. We get 
\begin{eqnarray}
\partial_i\tilde B^{i+} &=& 0 \label{bi}
\\
\epsilon^{ijkl}\partial_kA_l^+ &=& 0. \label{ai}
\end{eqnarray}
The above equations can be solved in terms of two 4D, $X$-depending potentials $\Lambda(X)$ and $\zeta_{ij}(X)$, 
with canonical dimensions zero and two, respectively:
\begin{eqnarray}
%\partial_i\tilde B^{i+}=0 &\Rightarrow& 
\tilde B^{i+}&=&\epsilon^{ijkl}\partial_j\zeta_{kl}
\label{zetadef}\\
%\epsilon^{ijkl}\partial_k A_l^+=0 &\Rightarrow& 
A_l^+&=&\partial_l \Lambda.
\label{lambdadef}
\end{eqnarray}
The definitions \eqref{zetadef} and \eqref{lambdadef} are invariant under the following gauge symmetry:
\begin{eqnarray}
\delta \Lambda&=&const \label{gauge1}
\\
\delta \zeta_{ij}&=&\partial_i \theta_j - \partial_j \theta_i, \label{gauge2}
\end{eqnarray}
where $\theta_i(X)$ is a local gauge parameter, signal of an electromagnetic-like structure on the boundary.

The boundary condition \eqref{bbordo}, written in terms of the potentials  $\Lambda(X)$ and $\zeta_{ij}(X)$, reads:
\begin{equation}
\epsilon^{ijkl}\partial_j\zeta_{kl}=\left(-\frac{a_4}{2}m^2-a_3\Box   \right)\partial^i\Lambda. \label{dual1}
\end{equation}
Notice that the coefficient $a_2$ is ruled out. Let us consider, in \eqref{dual1}, the term 
$a_3\Box  \partial^i\Lambda$, which refers to the term $\frac{a_3}{4}G^{ij}G_{ij}$ in the boundary Lagrangian \eqref{lbd}. If we apply the operator $\partial_i$ to \eqref{dual1}, we obtain:
\begin{equation}
0=\left(\frac{a_4}{2}m^2+a_3\Box  \right) \Box\Lambda,\label{dual2}
\end{equation}
which can be interpreted as a kind of generalized Klein-Gordon equation for the field $\Lambda(X)$ on the boundary, which leads to the momentum space propagator:
\begin{equation}
\label{propppp}
\Delta=-\frac{2}{a_4 m^2}\left(\frac{1}{p^2}-\frac{a_3}{a_3p^2-\frac{1}{2}a_4m^2}\right).
\end{equation}
Let us analyze in detail the form of this propagator which, as it stands, depends on two arbitrary parameters $a_3$ and $a_4$ (a third parameter $a_2$, as we have seen, although appearing in the boundary  Lagrangian ${\cal L}_{BD}$ \eqref{lbd}, does not affect the physics on the boundary). We first remark that $a_4$ should be different from zero: 
\begin{equation}
a_4\neq 0.
\label{a4notzero}\end{equation}
Otherwise, the field equation \eqref{dual2}, depending only on the $a_3$-term, would imply a badly IR divergent $\sim\frac{1}{p^4}$ propagator for the field $\Lambda(X)$. 

Stated that $a_4\neq0$, and recalling that our boundary is the flat Minkowski 4D spacetime with signature $(-1,1,1,1)$, we observe that it must be
\begin{equation}
a_3a_4\geq0,
\label{concordi}\end{equation}
$i.e.$, $a_3$ and $a_4$ must be concordant. This is to avoid a negative pole in the propagator \eqref{propppp}, which would correspond to a negative, unphysical, mass

The next step is to notice that 
\begin{equation}
a_4>0\ \ \ (\Rightarrow a_3\geq0).
\label{positivea4}\end{equation}
This is due to the fact that the leading term of the energy density is
\begin{equation}
T^{00}=a_4A_iA^i+O(\partial),
\label{energy}\end{equation}
where $T^{00}$ is the time-time component of the stress-energy tensor of the theory described by the boundary Lagrangian ${\cal L}_{BD}$ \eqref{lbd}.

Now we observe that the second of the two terms forming the propagator \eqref{dual2} has the wrong sign. It would correspond to a massive scalar particle with the wrong sign in the kinetic term, and therefore it would be something unphysical, like a kind of scalar ghost\footnote{In a different context, this is the method proposed by Pauli and Villars in \cite{Pauli:1949zm} to cure the U.V. divergencies in QFT.}. The only way out is to set $a_3$ equal to zero:
\begin{equation}
a_3=0.
\label{a3zero}\end{equation}

Once established that $a_4$ must be strictly positive, we can rescale the fields as follows
\begin{equation}
 \Lambda\rightarrow\frac{\Lambda}{\sqrt{a_4m^2}}\ \ ;\ \
\zeta\rightarrow \sqrt{a_4m^2}\zeta,
\label{rescaling}\end{equation} 
with the outcome that the boundary condition \eqref{dual1} becomes
\begin{equation}
\epsilon^{ijkl}\partial_j\zeta_{kl}=-\frac{1}{2}\partial^i \Lambda. \label{duality}
\end{equation}

We remark that the 4D equation \eqref{duality} closely resembles the analogous relation holding in the 3D case \cite{Amoretti:2012kb}. There, that equation represents the duality relation discussed in \cite{Aratyn:1984jz} which allows to interpret the boundary degrees of freedom as fermionic, although starting from a purely bosonic bulk theory. The fact that we are here recovering the same relation in 4D strongly suggest that the same mechanism of ``fermionization'' might occur also in 4D, as the consequence of the boundary condition in the dimensional reduction 5D$\rightarrow$ 4D {\it\`a\ la} Symanzik of a topological QFT. This nice result deserves further analysis.

%*********************************************************************************************
\section{4D Lagrangian}
%*********************************************************************************************

Differentiating the integrated Ward identity \eqref{w1} with respect to $J^j (x')$, with $x'$ lying on $x_4=0^+$, and then setting the sources to zero, we obtain the following relation:
\begin{equation}
\int{dx_4\,\delta^i_j\partial_i\delta^{(5)}(x-x')}=
\left.-i\partial_i \Delta_{A_j \tilde B^{i}}(x',x)\right|_{x_4=x'_4=0^+},\label{kaa}
\end{equation}
where $\Delta_{A_j \tilde B^{i}}(x',x)$ is the propagator defined as:
\begin{equation}
\Delta_{A_j \tilde B^{i}}(x',x)
\equiv 
\left. i\frac{\delta^2 \Gamma_c}{\delta J^j(x')\delta\tilde{J}_i(x)}\right|_{J=\tilde{J}=0},
\end{equation}
or, expressed in terms of the T-ordered product:
\begin{multline}
\left.\Delta_{A_j \tilde B^{i}}(x',x)\right|_{x_4=x'_4=0^+}=i\left<T\left(A_j^+(X')\tilde B^{i+}(X)\right)\right>=
\\
=i\theta(t-t')\left<\tilde B^{i+}(X)A_j^+(X')\right>+i\theta(t'-t)\left<A_j^+(X')\tilde B^{i+}(X)\right>.\label{prop}
\end{multline}
Substituting the equation \eqref{prop} in \eqref{kaa}, we obtain:
\begin{multline}
\delta^i_j\partial_i\delta^{(4)}(X-X')=-i\delta(t-t')\left<\left[\tilde B^{0+}(X),A_j^+(X')\right]\right>+
\\
-i\theta(t-t')\left<\partial_i \tilde B^{i+}(X)A_j^+(X')\right>-i\theta(t'-t)\left<A_j^+(X')\partial_i\tilde B^{i+}(X)\right>.\label{qwe}
\end{multline}
The second and the third term in the r.h.s of \eqref{qwe} vanish because of \eqref{bi}. Factorizing the delta function we obtain the following commutation relation for $\alpha=1,2,3$:\begin{equation}
\left[\tilde B^{0+}(X), A_\alpha^+(X')\right]_{t=t'}=-i\partial_\alpha\delta^{(3)}(X'-X).\label{cm1}
\end{equation}
In a similar way, we find two additional commutation relations for the fields $\tilde{B}^{0+}$ and $A^{\alpha+}$:
\begin{eqnarray}
\label{cm3}
\left[\tilde B^{0+}(X), \tilde B_j^+(X')\right]_{t=t'}&=&0,\label{cm31}\\
\left[A^{\alpha+}(X), A_\beta^+(X')\right]_{t=t'}&=&0.\label{cm32}
\end{eqnarray}
In conclusion, the commutation relations \eqref{cm1} , \eqref{cm31}  and \eqref{cm32} form the following algebra on the boundary:
\begin{eqnarray}
\left[\tilde B^{0+}(X), A_\alpha^+(X')\right]_{t=t'}&=&-i\partial_\alpha\delta^{(3)}(X'-X)
\label{algebra1}\\
\left[\tilde B^{0+}(X), \tilde B_j^+(X')\right]_{t=t'}&=&0\label{algebra2}\\
\left[A^{\alpha+(X)}, A_\beta^+(X')\right]_{t=t'}&=&0.\label{algebra3}
\end{eqnarray}
We can write the commutation relation \eqref{algebra1} in terms of the potentials $\Lambda(X)$ and $\zeta_{ij}(X)$. We find (remember that the index $\alpha$ runs over the spatial coordinates of the 4D boundary):
\begin{eqnarray}
\left[\Lambda(X'), \epsilon^{\alpha\beta\gamma}\partial_\alpha\zeta_{\beta\gamma}(X)\right]_{t=t'}
&=&
 -i\delta^{(3)}(X'-X)\label{cc1}
\\
\left[ \epsilon^{\alpha\beta\gamma}\zeta_{\beta\gamma}(X), \partial'_{\delta}\Lambda(X'),\right]_{t=t'}
&=&
 -i\delta_\delta^\alpha\delta^{(3)}(X'-X),\label{cc2}
\end{eqnarray}
The commutation relation \eqref{cc1} and \eqref{cc2} allow us to interpret the field $\Pi_{(\Lambda)} \equiv - \epsilon^{\alpha\beta\gamma}\partial_\alpha\zeta_{\beta\gamma}(X)$ and $\Pi_{(\tilde{\zeta}^{\alpha})} \equiv - \partial_{\delta}\Lambda(X)$ as the conjugate momenta of the fields $\Lambda$ and $\tilde \zeta_{\alpha} \equiv \epsilon_{\alpha\beta\gamma} \zeta^{\beta\gamma}$, respectively.

We are now ready to construct the boundary Lagrangian, which is the most general Lagrangian for the dynamical quantities $\{(\Lambda, \Pi_{(\Lambda)}),(\tilde \zeta_{\alpha}, \Pi_{(\tilde{\zeta}^{\alpha})}) \}$, gauge invariant according to \eqref{gauge1} and \eqref{gauge2}, whose EOM are compatible with the 4D duality relation \eqref{duality}. 

Some care is required to get the 4D Lagrangian from the commutation relations \eqref{cc1} and \eqref{cc2}, because we are dealing with a constrained system. Following the prescriptions described in \cite{Faddeev:1988qp},  we find that the most general 
boundary Lagrangian is:
\begin{equation}
\label{lagb}
\mathcal L=
\frac{1}{2} \epsilon^{\alpha\beta\gamma}\partial_{\alpha}\zeta_{\beta\gamma}\partial_t\Lambda + 
\frac{1}{2}\partial_\alpha\Lambda\epsilon^{\alpha \beta \gamma}\partial_t\zeta_{\beta\gamma}-
(\epsilon^{\alpha\beta\gamma}\partial_{\alpha}\zeta_{\beta\gamma})^2-\frac{1}{4}(\partial_\alpha\Lambda)^2.
\end{equation}
We stress that the 4D Lagrangian \eqref{lagb} is uniquely determined by the gauge symmetry \eqref{gauge1} and \eqref{gauge2} and the duality relation \eqref{duality}. Finally, notice that this Lagrangian cannot be trivially derived by substituting the expression for the fields $A^+_{i}$ and $\tilde B^{+}_{i}$ expressed in terms of the potential $\Lambda(X)$ and $\zeta_{ij}(X)$ in the boundary Lagrangian \eqref{lbd}, used to implement the Symanzik method. In fact, the first two terms in \eqref{lagb}, which correspond to the term $\Pi \dot{\Phi}$ of the Lagrangian $\mathcal{L}=\sum_i \Pi_i \dot{\Phi_i}-H$, break the covariance on the boundary, while the boundary Lagrangian \eqref{bbordo} is covariant in the manifold $x_4=0$.

%*********************************************************************************************
\section{Summary of results}
%*********************************************************************************************

In this paper, we considered the 5D topological BF theory 
\begin{equation}
\mathcal S_{BF}=\frac{1}{2}\int_M{d^5x\,\epsilon^{\mu\nu\rho\sigma\tau}F_{\mu\nu}B_{\rho\sigma\tau}}
\label{conclbf}
\end{equation}
in the axial gauge,  as the generating bulk model for a theory defined on the 4D boundary $x_4=0$. Only basic principles of QFT, like locality and power counting, have been used, following the main idea of ``separability'' introduced by Symanzik in \cite{Symanzik:1981wd}, and modified to avoid the explicit computation of the propagators for the complete theory (bulk \& boundary) \cite{Amoretti:2012kb,Blasi:2010gw}. 

We found that the most general ``separating'' boundary term, leading to physical degrees of freedom is
\begin{equation}
{\cal L}_{BD}=\delta(x_4)[\frac{a_2}{4}F_{ij}F^{ij}-\frac{a_4}{2}m^2A_iA^i].
\label{conclbd}\end{equation}
While the first Maxwell term does not affect the boundary physics, the presence of the second massive term is necessary (and crucial), since $a_4>0$ to insure a positive definite energy density.

The residual broken gauge invariance on the boundary is described by the WI
\begin{eqnarray}
\int{dx_4\,\partial_i J^i} &=& - \partial_i\tilde B^{i+}\label{conclw1}
\\
\int{dx_4\,\epsilon^{ijkl}\partial_k \tilde{J}_l} &=& \epsilon^{ijkl}\partial_k A^+_l,\label{conclw2}
\end{eqnarray}
which, taken at vanishing sources, lead to two ``Maxwell-like'' equations, solved in terms of a ``magnetic'' tensor potential and an ``electric'' scalar potential, which are the true 4D boundary degrees of freedom:
\begin{eqnarray}
\partial_i\tilde B^{i+}=0 &\Rightarrow& 
\tilde B^{i+}=\epsilon^{ijkl}\partial_j\zeta_{kl}
\label{conclzetadef}\\
\epsilon^{ijkl}\partial_k A_l^+=0 &\Rightarrow& 
A_l^+=\partial_l \Lambda.
\label{concllambdadef}
\end{eqnarray}
We found that only one boundary condition for the fields $A^+$ and $\tilde{B}^+$  is compatible with the Symanzik's criterion of separability. In terms of the potentials \eqref{conclzetadef} and \eqref{concllambdadef}, it reads
\begin{equation}
\epsilon^{ijkl}\partial_j\zeta_{kl}=-\frac{a_4}{2}m^2\partial^i \Lambda. \label{conclduality}
\end{equation}
This is one of the most important results of this paper. This ``electromagnetic'' duality relation is the 4D extension of the analogous relation used in \cite{Aratyn:1984jz} to get fermions from bosons. We find it here as the boundary condition for a topological QFT with boundary, and, as it happens in the 3D case \cite{Amoretti:2012kb}, it allows to guess that $fermionic$ 4D degrees of freedom are present, although the bulk theory is purely bosonic. We stress that $a_4>0$: the presence of the mass term in \eqref{conclbd} is crucial, since we showed that otherwise no physical solution exists.  We believe that is a common feature of the topological BF models in any spacetime dimensions: once a boundary is introduced {\it \`a la} Symanzik, the consequent boundary conditions, written in terms of the potentials found by putting on-shell the broken WI, play the role of ``fermionization'' duality relations. 

The other main result presented in this paper is the 4D Lagrangian. The WI \eqref{conclw1} and \eqref{conclw2} lead to an algebra of conserved currents which, written in terms of potentials, are interpreted as canonical commutation relations. Contrarily to what is usually done, from the canonical commutation relations we find the ``generating'' 4D Lagrangian, which is unique, thanks to the gauge invariance and to the duality relation \eqref{conclduality}:
\begin{equation}
\label{concllagb}
\mathcal L=
\frac{1}{2} \epsilon^{\alpha\beta\gamma}\partial_{\alpha}\zeta_{\beta\gamma}\partial_t\Lambda + 
\frac{1}{2}\partial_\alpha\Lambda\epsilon^{\alpha \beta \gamma}\partial_t\zeta_{\beta\gamma}-
(\epsilon^{\alpha\beta\gamma}\partial_{\alpha}\zeta_{\beta\gamma})^2-\frac{1}{4}(\partial_\alpha\Lambda)^2.
\end{equation}

In order to clarify the previous formal arguments, we finally propose a physical interpretation of our model.\\
Due to gauge invariance there are two conserved currents in the bulk:
\begin{eqnarray}
&J^{\mu}=\frac{\delta \mathcal{L}}{\delta A_{\mu}}=\partial_{\nu}\frac{\delta \mathcal{L}}{\delta\partial_{\nu}A_{\mu}}=\epsilon^{\mu \nu \rho \sigma \tau} \partial_{\nu}B_{\rho \sigma \tau}\\
&S^{\mu  \nu \rho}=\frac{\delta \mathcal{L}}{\delta  B_{\mu \nu \rho}}=\epsilon^{\mu \nu \rho \sigma \tau} \partial_{\sigma} A_{\tau}.
\end{eqnarray}
Following the argument in \cite{Levkivskyi:2008we}, we consider the edge excitations as a deformation of the boundary caused by the bulk currents flowing towards the edge. We parametrize the deformation by $h(X)$ and we represent the edge currents $\overline{J}^i$ and $\overline{S}^{ijk}$ as:
\begin{eqnarray}
&\overline{J}^i=\int_{-x^4_0}^h dx^4 J^i\\
&\overline{S}^{ijk}= \int_{-x^4_0}^h dx^4S^{ijk},
\end{eqnarray} 
where $x^4_0$ is an auxiliary boundary, where the bulk and edge currents match. In the low energy limit, where $h$ is much smaller than the typical deformation wavelength, and keeping in mind our gauge choice (2.6) and (2.7), the previous integrals can be approximated by the following expressions (see Appendix D of \cite{Levkivskyi:2008we} for details):
\begin{eqnarray}
&\overline{J}^i=(h+x_0^4) \epsilon^{ijkl}B_{jkl}(X,0)\\
&\overline{S}^{ijk}=(h+x_0^4) \epsilon^{ijkl}A_l(X,0).
\end{eqnarray} 
Now, according to the equations (7.5) and (7.6), we have $A_l(X,0)=\partial_l \Lambda(X)$ and $B_{klm}(X,0)=\epsilon_{klmi}\epsilon^{ijku}\partial_j \zeta_{ku}(X)$, consequently, the duality relation $\epsilon^{ijkl} \partial_j \zeta_{kl}=-\frac{1}{2}\partial^i \Lambda$ ((5.16)) yields:
\begin{equation}
\overline{J}^i=-\frac{1}{2}\epsilon^{ijkl}\overline{S}_{jkl}.
\end{equation}
If we look at $\overline{J}^i$ and $\overline{S}^{ijk}$ as a generalized charge density current and spin density current respectively, the previous relation is exactly a higher dimensional generalization of the well known one  which occurs between the charge density current and the spin density current on the edge of a (3+1)D Topological Insulator \cite{zhang}.

%*********************************************************************************************

\end{document}